\documentclass[aps,prb,reprint,amsmath,amssymb,superscriptaddress,showpacs,floatfix]{revtex4-1}
\usepackage{graphicx,amsfonts,times,bm,amsmath,verbatim,color,array}
\usepackage{ifthen,braket,xcolor,bm}
\usepackage[colorlinks, allcolors=blue]{hyperref}
\usepackage{natbib}
\usepackage{braket}
\usepackage{cancel}                  

\newcommand{\ketn}[1]{ {#1} \rangle}
\newcommand{\ua}{\uparrow}
\newcommand{\da}{\downarrow}

\definecolor{lred}{RGB}{255,95,95}   

\begin{document}


\title{Fermion parity gap and exponential ground state degeneracy of the one-dimensional Fermi gas with intrinsic attractive interaction}


\author{Monalisa Singh Roy}
\email{monalisa12i@bose.res.in}
\author{Manoranjan Kumar}
\email{manoranjan.kumar@bose.res.in}
\affiliation{S. N. Bose National Centre for Basic Sciences, Block - JD, Sector - III, Salt Lake, Kolkata - 700106, India}



\author{Jay D. Sau}
\email{jaydsau@umd.edu}
\affiliation{Condensed Matter Theory Center and Joint Quantum Institute, Department of Physics, University of Maryland, College Park, Maryland 20742-4111, USA}

\author{Sumanta Tewari}
\email{stewari@g.clemson.edu }
\affiliation{Department of Physics and Astronomy, Clemson University, Clemson, SC 29634, USA}



\date{\today}

\begin{abstract}
We examine the properties of a one-dimensional (1D) Fermi gas with attractive intrinsic (Hubbard) interac- tions in the presence of spin-orbit coupling and Zeeman field by numerically computing the pair binding energy, excitation gap, and susceptibility to local perturbations using the density matrix renormalization group. Such a system can, in principle, be realized in a system of ultracold atoms confined in a 1D optical lattice. We note that, in the presence of spatial interfaces introduced by a smooth parabolic potential, the pair binding and excitation energy of the system decays exponentially with the system size, pointing to the existence of an exponential ground state degeneracy, and is consistent with recent works. However, the susceptibility of the ground state degeneracy of this number-conserving system to local impurities indicates that the energy gap vanishes as a power law with the system size in the presence of local perturbations. We compare this system with the more familiar system of an Ising antiferromagnet in the presence of a transverse field realized with Rydberg atoms and argue that the exponential splitting in the clean number-conserving 1D Fermi system is similar to a phase with only conventional order.
\end{abstract}

\pacs{}

\maketitle

\section{Introduction} \label{Introduction}
Topological quantum computation (TQC) promises the realization of robust fault-tolerant quantum information processing~\cite{kitaev2001unpaired,nayak2008non}, and tremendous experimental efforts are being made to find and fabricate systems supporting TQC~\cite{mourik2012signatures,deng2012anomalous,das2012zero,rokhinson2012fractional,churchill2013superconductor,
finck2013anomalous,albrecht2016exponential,deng2016majorana,zhang2017ballistic,chen2017experimental,nichele2017scaling,zhang2018quantized}. Among the most promising candidates for realizing TQC in condensed-matter systems are solid-state semiconductor thin films and nanowires, which are theoretically predicted to support a topological superconducting (TS) phase in the presence of a proximity induced superconducting pair potential $\Delta_{ind}$, Rashba spin-orbit coupling (SOC) $\alpha$, and an externally applied Zeeman field $h$, in the parameter space spanned by the weak coupling mean field equation $h^2>(\Delta_{ind}^2+\mu^2)$~\cite{sau2010generic,tewari2010theorem,alicea2010majorana,sau2010non,lutchyn2010majorana,oreg2010helical,
stanescu2011majorana}. The TS phase is defined by the emergence of mid-gap non-Abelian topological quasiparticles known as Majorana zero modes (MZMs) localized at the topological defects~\cite{read2000paired,kitaev2001unpaired,nayak2008non,beenakker2013search,elliott2015colloquium}.
Alongside the solid state systems, it has also been proposed that the system of ultracold fermions confined in optical lattice~\cite{bloch2005,lewenstein2007} in the presence of SOC, Zeeman field, and a mean-field $s$-wave superfluid pair potential supports a TS phase with MZMs as edge modes~\cite{Chuanwei_2008,Jiang,Gong,Seo_1,Seo_2,Seo_3}. Although effective SOC and Zeeman field can be experimentally generated in a one-dimensional (1D) system of ultracold atoms~\cite{Lin,Wang,Cheuk,Cui,Zhai}, the study of 1D fermions with intrinsic attractive interactions induced by Feshbach resonance~\cite{Fu} in the framework of mean-field theory is problematic. This is because the reduced dimensionality results in strong pair phase fluctuations, and true superfluid long-range order in one dimension is destroyed. It was shown by Sau {\it et al.}~\cite{Sau2011Halperin} and Fidkowski {\it et al.}~\cite{Fidowski2011} that in the presence of phase fluctuations, the ground state would differ from the conventional phase by the presence of an exponential ground state degeneracy in the absence of phase slips. Phase slips however, reduce ground state  degeneracy from exponential to power law. Later on it was clarified that such phase slips are likely to be rare in systems of ultracold atoms ~\cite{Ruhman_2017}. Regardless of the degeneracy, it was argued~\cite{Kane2017} that number conserving systems  are characterized by long-range order in a transverse-field Ising degree of freedom that is related to the original fermions by a Jordan-Wigner transformation~\cite{1934Jordan}.  However, the non-local order parameter in the relevant transverse-field Ising model relies on a decomposition of the system into fermions and bosons and as such is not obviously related to a non-local observable in the original fermion model.  TQC, however, relies most importantly on the exponential ground state degeneracy, which can be used to build non-local qubits and quantum gates immune to local perturbations. In this paper we will focus exclusively on the existence of exponential ground state degeneracy in a 1D Fermi gas with SOC, Zeeman field, and intrinsic attractive interactions in the presence of a smooth parabolic potential, as considered in Ref.~[\onlinecite{Erez_Parabolic}]. The case for topological superfluidity in such systems is made through the degeneracy of the entanglement spectrum and Majorana correlation function, provided there is a fermionic gap \cite{2008_Li, 2013_Kraus,2017_Iemini, 2018_Chen}. The present system [Eqs.~\eqref{eqH} and \eqref{eq.1: Hamiltonian Parabolic}], on the other hand, is gapless to fermionic excitations because of the phase slips \cite{Sau2011Halperin, Fidowski2011}. For this reason, and because we are exclusively focused on the existence of the exponential ground state degeneracy, which is the only property needed for TQC, we have not attempted to explore topological properties through the degeneracy of the entanglement spectrum or other indicators.

It was recently proposed~\cite{Erez_Parabolic} that the exponential ground state degeneracy of a 1D Fermi gas with SOC, Zeeman field, and intrinsic attractive interactions could be established through the use of a smooth parabolic potential \--- which spontaneously occurs in ultracold atom systems confined by a harmonic trap potential.
 The parabolic potential $V(r)$ introduces smooth interfaces between regions defined by $h > \mu_\text{\normalfont eff}$ and $h < \mu_\text{\normalfont eff}$, where $\mu_\text{\normalfont eff}$, defined by $\mu_\text{\normalfont eff} = \mu-V(r)$, represents the effective chemical potential. If the superconducting order could be treated in mean-field theory, as in the case of the spin-orbit coupled semiconductor-superconductor heterostructure with proximity-induced superconductivity and an externally applied Zeeman field ~\cite{sau2010generic,tewari2010theorem,alicea2010majorana,sau2010non,lutchyn2010majorana,oreg2010helical,
stanescu2011majorana}, these regions, in the limit of a small mean-field superconducting pair potential $\Delta_{ind}\rightarrow 0$,  would be the topological superconducting and ordinary superconducting phases, respectively, and for $h > \mu_\text{\normalfont eff}$ the system would have a two-fold exponential degeneracy of the ground state  ~\cite{sau2010generic,tewari2010theorem,alicea2010majorana,sau2010non,lutchyn2010majorana,oreg2010helical,
stanescu2011majorana}. 
 In the present system, however, we have intrinsic attractive interaction, and superconductivity cannot be treated in mean-field theory. Despite that, the ground state has been proposed \cite{Erez_Parabolic} to be doubly degenerate (up to an exponentially small splitting for finite-sized system) because of an exponential ground state degeneracy associated with the fermion parity of the system. Hence, an exponential decay of the excitation gap $\Delta$ and pair binding energy $E_b$ with increasing system size $N$ indicating the absence of a fermion parity gap could confirm the existence of exponential ground state degeneracy in the 1D Fermi gas with attractive interactions.
In this paper, we numerically study the 1D Fermi gas system proposed in Ref.~[\onlinecite{Erez_Parabolic}] to search for  exponential ground state degeneracy in the fermion parity gap and the excitation gap. The question of exponential splitting of the ground state degeneracy is at the heart of topological protection of qubits in topological quantum computation ~\cite{kitaev2001unpaired,nayak2008non}. As is clear from the Luttinger liquid (LL) analysis of a spin-orbit coupled Fermi gas with attractive interactions, Bosonization of a clean system leads to an exponential degeneracy in pairs of SOC gases~\cite{Terhal2012,Landau2016,Erez_Parabolic}. Within this formalism, power law splitting can be generated by back scattering induced phase slip terms shared between pairs of wires~\cite{Sau2011Halperin}. Microscopically, the back scattering originates (at weak interactions) from scattering between different Fermi surfaces~\cite{Giamarchi} and therefore requires breaking of momentum conservation by some impurity.

In this work, we show through numerical calculations that in the {\it clean} 1D Fermi gas with attractive Hubbard interaction, the pair binding energy and the excitation gap vary exponentially with the system size, indicating the vanishing of the fermion parity gap and a two-fold ground state degeneracy. However, by numerically studying the expectation values of relevant local operators and the effect of impurity potentials on the ground state degeneracy, we conclude that the ground state degeneracy of the number-conserving 1D Fermi gas ceases to be exponential in the presence of local perturbations. Thus, the main result of this work is the numerical demonstration that the fermion parity gap and the ground state degeneracy of the number-conserving 1D Fermi gas are no longer exponential in the presence of local perturbations and are thus not suitable for TQC. The proposal of creating  MZMs in a 1D spin-orbit coupled interacting Fermi gas appeared in Ref.~[\onlinecite{Erez_Parabolic}] based on theoretical calculations, and a rigorous numerical verification needs to be made for such novel propositions before experiments are designed on the basis of the same. We have employed the density matrix renormalization group (DMRG), a state-of-the-art numerical technique for low-dimensional systems capable of calculating accurate results away from half-filling of electrons and for systems containing complex interactions such as SOC, to probe the important theoretical predictions in Ref.~[\onlinecite{Erez_Parabolic}]. These numerical results are thus important both in the context of designing experiments for realizing TQC, and for addressing fundamental questions regarding the existence of exponential ground state degeneracy in 1D Fermi gas systems with intrinsic attractive interactions.

This paper is organized in four sections. In Sec. \ref{Model and Method}, we introduce the model and the numerical technique and also discuss the main criteria used for identifying exponential ground state degeneracy in the system. In Sec. \ref{Results}, we first analyze the energy gaps in the clean system to find evidence of exponential ground state degeneracy, if any. Then, we examine the system for indistinguishability in local operator measurements for the two exponentially degenerate states and study the robustness of the energy degeneracy in the presence of local impurities. Finally, we compare these results with that of the transverse field Ising model to argue whether the exponential ground state degeneracy apparent in the clean system truly reflects an underlying topological phase, as claimed in Ref. ~[\onlinecite{Erez_Parabolic}] . We conclude with a brief discussion of the reported results and their possible impact on the current understanding of topological properties in 1D ultracold systems in Sec. \ref{Discussion}.

\section{Model and Method} \label{Model and Method}
We consider a 1D spin-orbit coupled Fermi gas with an attractive $s$-wave interaction with
strength $g$ driven by a Feshbach resonance. The SOC together with
Zeeman coupling was experimentally realized~\cite{Lin} in gases of ultra-cold atoms through
the application of a pair of Raman lasers with recoil wave vector $k_r$.
The lasers couple to two hyperfine atomic states represented by the pseudospins
$\sigma=\vert \ketn{\ua},\vert \ketn{\da}$ (e.g., $\vert \ketn{\ua} = \vert \ketn{f=9/2,m_F=-7/2}$ and
$\vert \ketn{\da} = \vert \ketn{f=9/2,m_F=-9/2}$), as has been observed in
experiments on $K^{40}$
gases~\cite{Williams}, through a third state.
The Raman coupling together with the incident Feshbach resonance~\cite{Williams}
leads to an effective Hamiltonian~\cite{Lin} for the Fermi gas of atoms
that is written as
\begin{align}
H= & \int dx \left[ \sum_{\alpha,\beta}  \psi^\dagger_\alpha(x)[-\frac{1}{2m}\partial_x^2  \delta_{\alpha\beta}+i\zeta\partial_x\sigma^{(x)}_{\alpha\beta}+\Omega\sigma^{(z)}_{\alpha\beta}]\psi_\beta(x) \right. \nonumber\\
& \left. -g\psi^\dagger_{\ua}(x)\psi^\dagger_{\da}(x)\psi_{\da}(x)\psi_{\ua}(x) \right],\label{eqH}
\end{align}
where $\zeta=k_r/m$ is the strength of the SOC and $\Omega$ is
the strength of the Raman coupling. Here $\psi^\dagger_\sigma(x)$ is the creation
operator for atoms of mass $m$ with pseudospin $\sigma$ at position $x$,
 and we have chosen units so that $\hbar=1$. $\sigma^{(a)}_{\alpha,\beta}$
represent the standard Pauli matrices with $a=x,y,z$, and $\alpha, \beta$ are the spin indices.

For the purposes of numerical calculation, it is necessary to discretize this Hamiltonian with a lattice parameter $a$ that is much smaller than the inverse density $\nu^{-1}$, where $\nu=n/2N$ represents the filling fraction of the system of size $N$, containing $n$ electrons. Within this approximation, the above continuum Hamiltonian can be written as the sum of three contributions, i.e., an on-site interaction $H_{\text{U}}$, uniform Zeeman field $H_{\text{Z}}$ and  Rashba spin-orbit interaction $H_{\text{SOC}}$. In addition, a parabolic potential with tunable parameter $k^{\prime}$ controls the electron density profile $H_{\text{para}}$.
The model Hamiltonian for the system can be written as
\begin{equation} \label{eq.1: Hamiltonian Parabolic}
 H = H_{\text{\normalfont t}} + H_{\text{\normalfont U}} + H_{\text{\normalfont SOC}} + H_{\text{\normalfont Z}} + H_{\text{\normalfont para}},
\end{equation}
where
\begin{align*}
 H_{\text{\normalfont t}} = & -t \sum_{i, \sigma} \left( C_{i, \sigma}^{\dagger} C_{{i+1, \sigma}} + h.c. \right)\text{\normalfont,} \hspace{1ex} H_{\text{\normalfont U}} =  U \sum_i  n_{i, \uparrow} n_{i, \downarrow}\text{\normalfont,} \\ \nonumber
 H_{\text{\normalfont SOC}} = & + i \alpha \sum_{i} \left( C_{i, \uparrow}^{\dagger} C_{i+1, \downarrow}  + C_{i, \downarrow}^{\dagger} C_{i+1, \uparrow} - h.c. \right)\text{\normalfont,} \\ \nonumber
H_{\text{\normalfont Z}} = & \hspace{2ex} h \sum_{i} S_{i}^{z}\text{\normalfont,} \hspace{1ex} H_{\text{\normalfont para}} = \sum_i \left( \dfrac{1}{2} k^{\prime} r^2\right) (n_{i, \uparrow} + n_{i ,\downarrow})\text{\normalfont.} \\ \nonumber
\end{align*}
To match the continuum Hamiltonian in Eq.~\eqref{eqH}, in Eq.~\eqref{eq.1: Hamiltonian Parabolic} we set the nearest-neighbor hopping to be $t=\frac{1}{2 m a^2}$, the SOC strength to be $\alpha=\zeta/2 a$, the Zeeman coupling as $h=\Omega$, and the on-site Hubbard potential as $U=-g/a$.
To simplify the presentation of results, we choose energy units for our calculation
so that the hopping amplitude $t$ is set to unity.
 The parabolic potential tuning parameters are $k^{\prime}=k/N^{2}$ and $r=(N+1)/2 - i$, where $i$ and $N$ are site indices and system size, respectively. We study a low filling fraction of the electrons ($\nu=0.10$ and $0.20$), and focus on the attractive interaction regime $U \in [-1.00,-4.00]$. {A given linear density $\rho$ of fermions in the continuum corresponds to a filling fraction $\nu=\rho a$, which becomes vanishingly small
in the true continuum limit $a\rightarrow 0$ relevant to experiments on Fermi gas~\cite{Lin}. We expect the low filling fractions $\nu=0.10$ to be small enough to match the continuum limit. It should be noted that in order to obtain a large enough superconducting gap we chose $|U/t| \geq 1$ in our calculations, which corresponds to a 1D scattering length $a_{sc}=\frac{2}{mg} = \frac{4ta}{\vert U \vert} \leq 4 a$. Therefore, while $|U/t|\propto a \rightarrow 0$ in the continuum limit, the value $U/t=-1$ corresponds to a scattering length of $\sim 4a$ that significantly exceeds the discretization spacing $a$ so that we can still expect continuum behavior in this limit. Ideally, verification of convergence in the continuum limit requires us to choose different values of $a$ while keeping the physical variables $\rho, g, m, \zeta$, and $\Omega$ in Eq.~\eqref{eqH} fixed. Unfortunately, this calculation would require going to much larger values of system size $N$, which is beyond the scope of this work. Thus, while the goal is to study the continuum Hamiltonian, it is difficult to get to the true continuum limit ($a\rightarrow 0$) because of the complicated Hamiltonian, and $|U/t| \ge 1$ is chosen because for smaller $|U|$ it would be difficult to observe anything of substance for large system size $N$.
}

We have used the DMRG~\cite{DMRG1,DMRG2,DMRG3,DMRG4} method, a state-of-the-art numerical technique for calculating the eigenvalues and eigenvectors of low-dimensional systems, for solving the Hamiltonian in Eq.~\eqref{eq.1: Hamiltonian Parabolic}. In the fermionic system under study, the spin degrees of freedom are not conserved, and hence, the Hamiltonian dimension is significantly large. The eigenvectors of the density matrix corresponding to $m \simeq 700$ largest eigenvalues have been retained to maintain a reliable accuracy. More than ten finite DMRG sweeps have been performed for each calculation so that the error in calculated energies is less than 1\%.

We study two spectral characteristics of the system: the vanishing pair binding energy ($E_b$) or the parity gap and an exponential decay of the excitation energy gap $\Delta$, defined as,
\begin{subequations}
\begin{align}
E_b (n,N) & = \dfrac{1}{2} \left[ E_{0}(n+1,N) + E_{0}(n-1,N) - 2 E_{0}(n,N) \right],\\
\Delta (n,N) &=E_{1} (n,N) - E_{0} (n,N).   \label{eq.2: Excitation Gap}
\end{align}
\end{subequations}
$E_{0}(n,N)$ and $E_{1}(n,N)$ are the ground state energy and the first excited state energy with $n$ electrons in a system of size $N$. In the absence of $U$ and $\alpha$, the spin up and spin down electronic bands split in the presence of $h$. But to create intra-band pairing correlations, an attractive $U$ is needed. Now the SOC interactions applied along the $x$ direction generate a momentum-dependent magnetic field along the $x$ axis.

\section{Results} \label{Results}

In this section we present the numerical studies investigating the existence of a robust exponential degeneracy of the ground state in 1D ultracold atoms in the presence of intrinsic interactions, SOC, Zeeman field, and only a parabolic potential. We first study the scaling of the binding energy $E_b$ and the lowest excitation gap $\Delta$ with system size $N$, the exponential decay of which is considered a hallmark of edge modes and a resulting utility in TQC. Next, we examine the local indistinguishability criteria for this system corresponding to a local charge density operator and also check the robustness of energy gap degeneracy against local impurities in wire. Finally, we compare the observed results with that of the transverse field Ising model and discuss whether our system is truly topological notwithstanding the exponentially decaying energy gaps in the clean system.

\subsection{Energy gaps} \label{sub.Energy Gaps}
	
\begin{figure}[t]  \label{Eb:U_diff}
\includegraphics[width=0.95\linewidth]{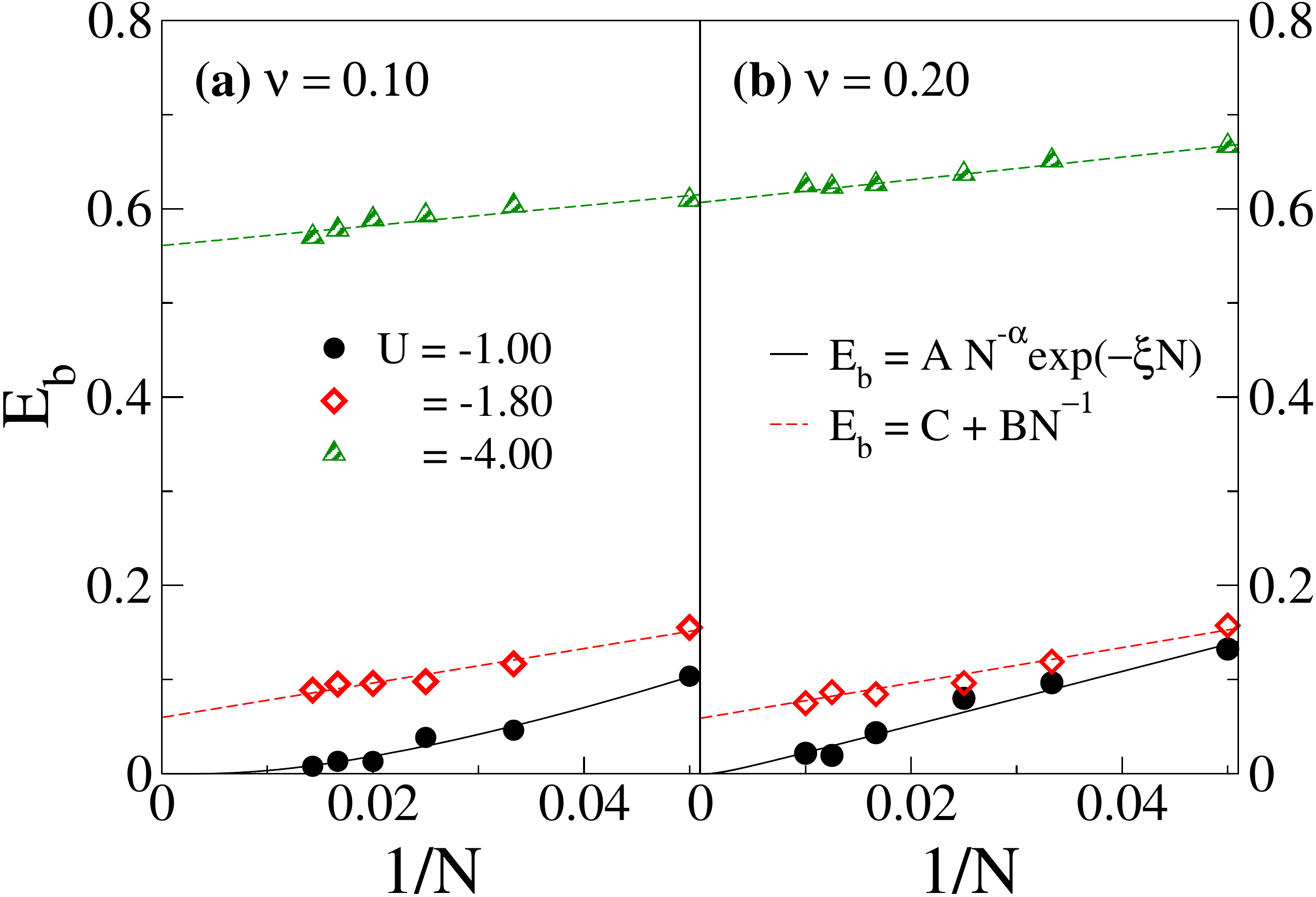}
\caption{\small (Color online) Variation of the pair binding energy $E_b$ with $1/N$ for different $U$ at $\alpha=0.20$, $h=0.40$, $k=3$, for (a) $\nu=0.10$, and (b) $\nu=0.20$. The dashed curves represent power law fitting with parameters $(C,B)$. In (a), $(C,B)$ are extracted as $(0.064,1.23)$ and $(0.57,0.58)$ for $U=-1.80$ and $U=-4.00$, respectively. In (b), $(C,B)$ are extracted as $(0.059,1.88)$ and $(0.61,1.20)$ for $U=-1.80$ and $U=-4.00$, respectively. The solid black curve represents a vanishing exponential with fitting parameters $(A,\alpha,\xi)$ extracted corresponding to $U=-1.00$ in (a) as $(8.76,1.39,0.014)$ and in (b) as $(2.90,1.00,0.0025)$.}  \label{fig.2}
\end{figure}

The non-topological conventional phase is expected to be adiabatically connected to
the conventional $s$-wave superconductor with Cooper pairs as the only low-energy degrees of freedom. The phase with exponential ground state degeneracy, on the other hand, is expected to harbor low-energy fermionic edge modes, so that the fermion parity of the system is no longer gapped. We start by numerically searching for exponential ground state degeneracy in the system by studying the size dependence of the parity gap  $E_b$. We have shown the variation of $E_b$ with $N$ for different attractive Hubbard potentials $U$, at electron fillings $\nu=0.10$ in Fig.~\ref{fig.2}(a) and at $\nu=0.20$ in Fig.~\ref{fig.2}(b). We find that the parity gap $E_b$, for the stronger $U=-4.00$ and $U=-1.80$, saturate to a finite intercept as $N$ increases in both Figs.~\ref{fig.2}(a) and~\ref{fig.2}(b). This is consistent with conventional Cooper pairing expected for the non-topological phase. In contrast, the parity gap $E_b$ is seen to vanish in the thermodynamic limit for a weakly attractive potential, $U=-1.00$ for both $\nu=0.10$ and $\nu=0.20$, suggesting a phase that is qualitatively distinct from the conventional phase at $U=-4.00$ and $U =-1.80$, in these limits. Here parabolic potential with $k=3$ has been kept fixed, and moderate Zeeman field ($h=0.40$) and SOC strength ($\alpha=0.20$) have been used.

\begin{figure}[t]  \label{Delta:U_diff}
\includegraphics[width=0.95\linewidth]{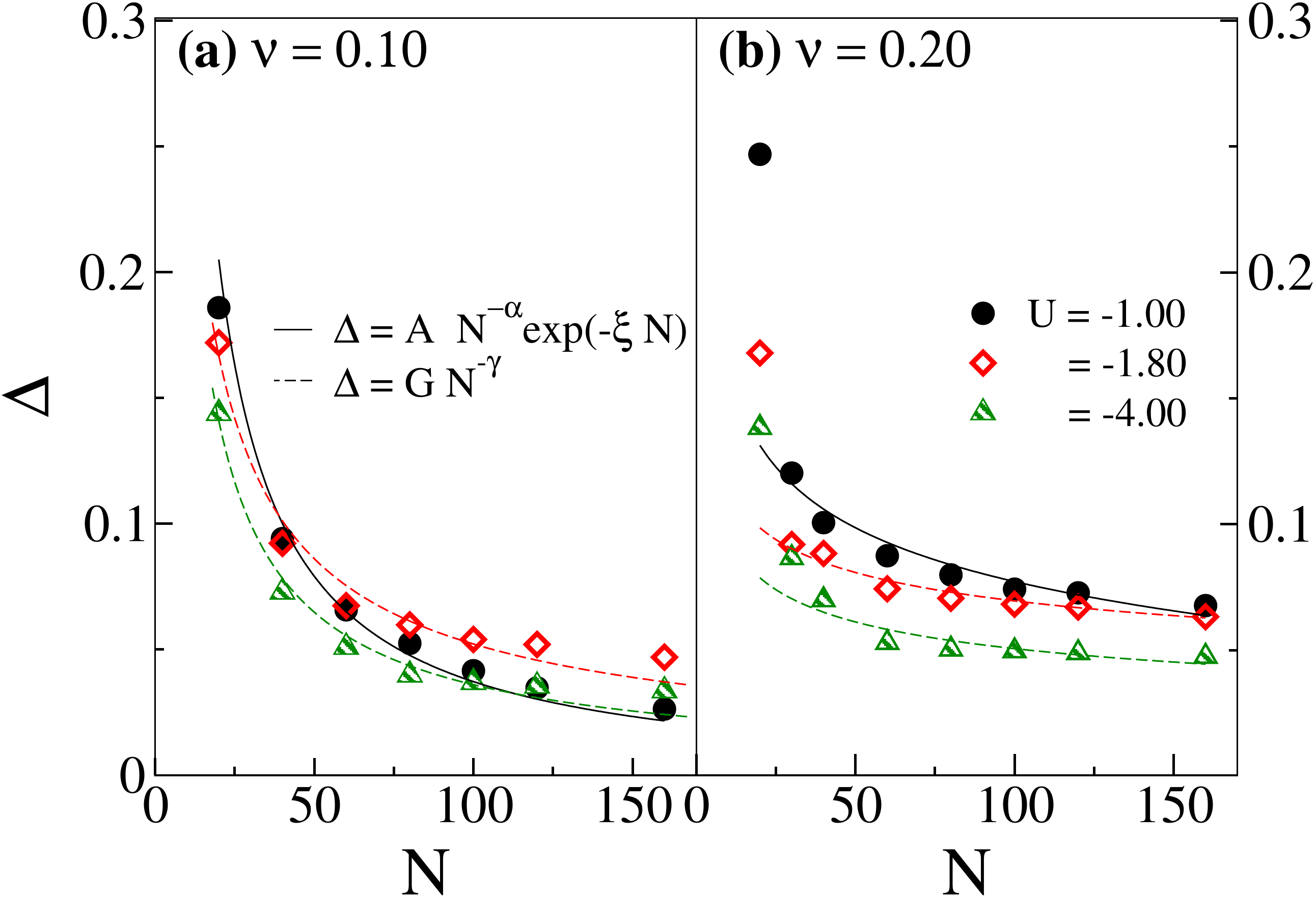}
\caption{\small (Color online) Variation of $\Delta$ with $N$ for different $U$ at $\alpha=0.20$, $h=0.40$, $k=3$, for (a) $\nu=0.10$, and (b) $\nu=0.20$. The dashed curves represent power law fitting with parameters $(G,\gamma)$. In (a), $(G,\gamma)$ are extracted as $(1.45,0.72)$ and $(1.78,0.85)$, for $U=-1.80$ and $U=-4.00$, respectively. In (b), $(G,\gamma)$ are extracted as $(0.19,0.22)$ and $(0.18,0.27)$, for $U=-1.80$ and $U=-4.00$, respectively. The solid curve represents an exponential fitting with parameters $(A,\alpha,\xi)$ extracted for $U=-1.00$ in (a) as $(3.07,0.94,0.001)$, and in (b) as $(0.31,0.28,0.001)$.} \label{fig.3}
\end{figure}

Next, in Fig.~\ref{fig.3} we show the variation of the excitation gap $\Delta$ with $N$ for different attractive potentials, $U=-1.00$, $-1.80$, and $-4.00$ at filling $\nu=0.10$ and $0.20$. All the other parameters have been kept the same as in Fig.~\ref{fig.2}. First, we note that in Fig.~\ref{fig.3}(a), the power law decrease of the excitation energy with system size $N$ in the conventional non-topological  phase for $U = -4.00, -1.80$ is consistent with the excitations arising from phonon modes. By contrast, the energy gap $\Delta$ for $U=-1.00$ fits better with an exponential dependence on the system size than a pure power law, as indicated by the variance of the fitted curve with the data. The variance  corresponding to the exponential fitting is at least ten times smaller than that for the pure power law fitting. This, apart from possible finite-size errors, is qualitatively consistent with the behavior of the pair binding energy $E_b$ shown in Fig.~\ref{fig.2}, demonstrating the possible existence of an exponential ground state degeneracy for a weakly attractive potential, $U=-1.00$. These results are also consistent with the theoretical predictions of the existence of exponential degeneracy, expected of a TS phase in a spin-orbit-coupled Fermi gas in the presence of a parabolic trap potential~\cite{Erez_Parabolic}. {Similar behavior of $\Delta$ is found at higher filling fraction $\nu=0.20$ as well, as $\Delta$ shows exponential decay as a function of $N$ for $U=-1.00$, and shows a pure power law decrease with $N$ for $U=-1.80$ and $-4.00$ in Fig.~\ref{fig.3}(b).}



\subsection{Local indistinguishability and robustness} \label{sub.Local Indistinguishability and Robustness}

\begin{figure}[t]  \label{fig.Local_operator}
\includegraphics[width=0.92\linewidth]{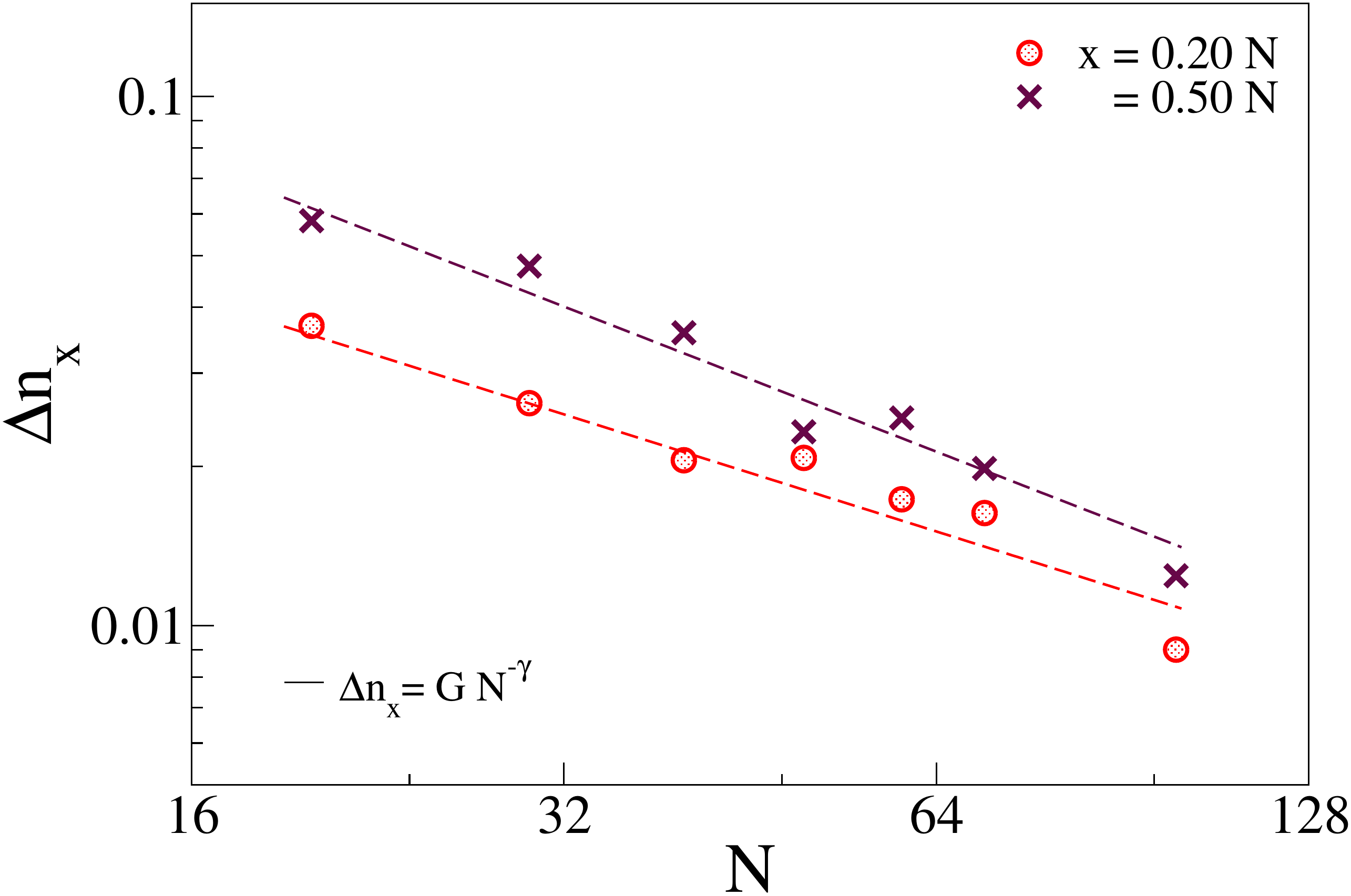}
\caption{\small (Color online) Log-log plot showing variation of $\Delta n_x$ with $N$; $\Delta n_x$ represents the difference in local charge density between the lowest excited state $\vert 1 \rangle$ and the ground state $\vert 0 \rangle$, at a local position ($x$) for $U=-1.00$, $\nu=0.10$, $\alpha=0.20$, $h=0.40$, and $k=3$. The power law fitting parameters $(G,\gamma)$, for $x=0.2N$ and $0.5N$ are $(0.33,0.73)$ and $(0.94,0.91)$, respectively.} \label{fig4}
\end{figure}

To be useful for topological qubits in TQC the lowest-lying states in gapped phases should be gapped by an exponentially small splitting which is robust to local perturbations. This is related to another indicator of such phases, namely, the absence of an order parameter, or equivalently, local indistinguishability of the pair of exponentially degenerate states~\cite{nayak2008non}. To determine whether the two exponentially degenerate ground states are locally distinguishable, we study a local operator defined as, $\Delta n_x = \langle n_{1} \rangle_x - \langle n_{0} \rangle_x$. $\Delta n_x$ represents the difference in local charge density between the lowest excited state $\vert 1 \rangle$ and the ground state $\vert 0 \rangle$, at a local position ($x$) of the system. The averaged difference in the charge density $\sum_{x=1}^{N} \Delta n_x $ taken over the entire system, which is a global operator, vanishes for any arbitrary $N$, since this is a charge-conserving system. However, we find that local measurements at, say, $x=0.2N$ and $0.5N$, show a power law dependence of $\Delta n_{x}$ on $N$ (Fig.~\ref{fig4}) for system sizes studied up to $N=120$.
Power law variation of $\Delta n_x$ with $N$ is observed for any other $x$ on the 1D wire too. This is in contrast to an exponential decay of $\Delta n_x$ as expected from a system with topological order and exponential ground state degeneracy.
This observation suggests that the apparent exponential degeneracy of the number-conserving spin-orbit coupled 1D Fermi gas is possibly different from what is expected in a topological phase.

The local indistinguishability of different ground states is intimately connected to the robustness of the exponential degeneracy of the ground states to local perturbations. To present this point more concretely, we consider impurity potentials at two sites in the bulk of the system, written as
\begin{equation}
H_{\text{\normalfont im}} = V_{\text{\normalfont im}} \left( n_{N/2} + n_{N/2 +1} \right)
\end{equation}

\begin{figure}[t]
\includegraphics[width=0.87\linewidth]{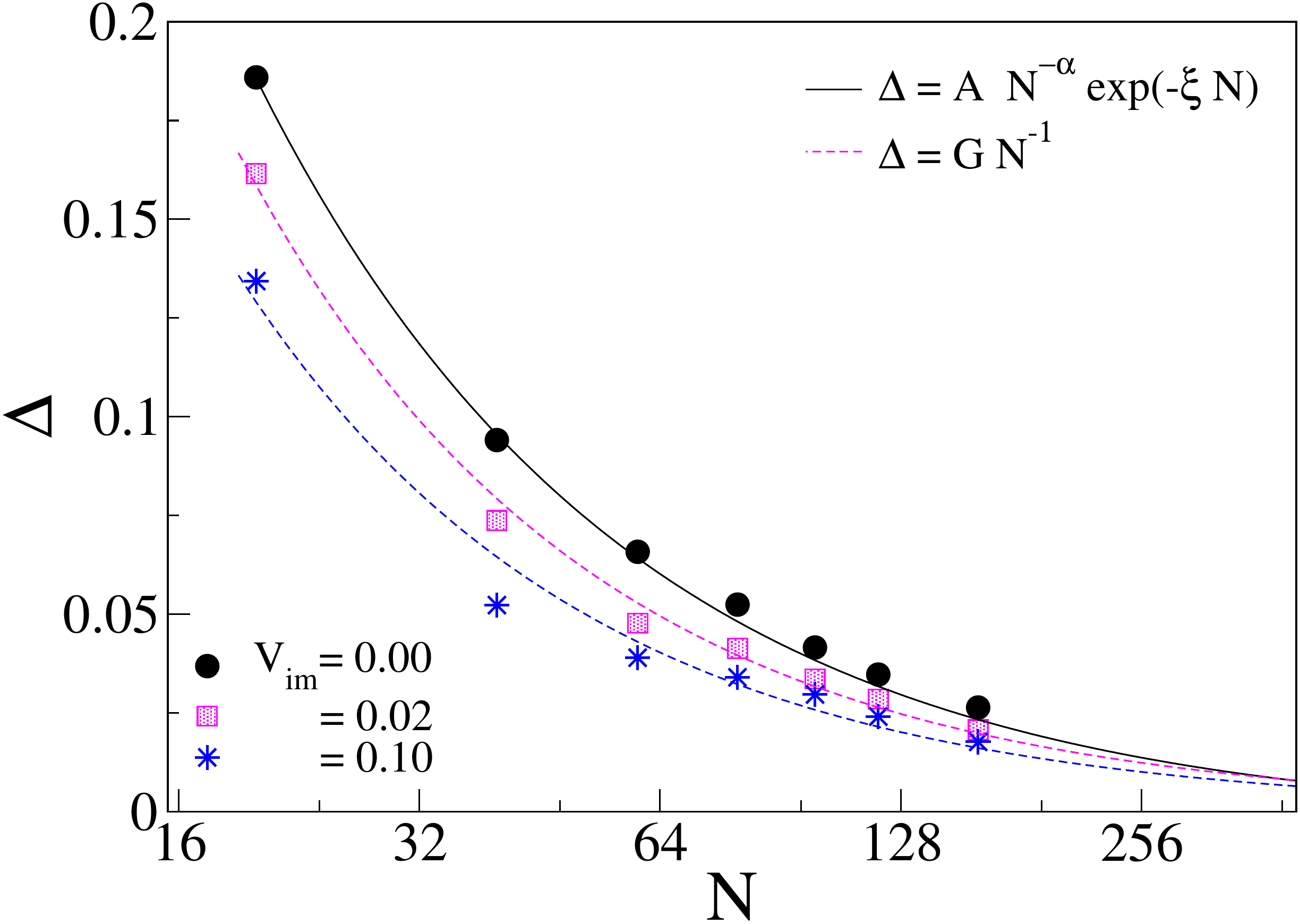} \label{fig.Robust_impurity}
\caption{\small (Color online) Semi-log plot showing variation of $\Delta$ with $N$ for different $V_{\text{\normalfont im}}$, at $U=-1.00$, $\nu=0.10$, $\alpha=0.20$, $h=0.40$ and $k=3$. The dashed curves represent power law fitting with the parameter $G = 3.17$ and $2.58$, for $V_{\text{\normalfont im}}=0.02$ and $0.10$, respectively. The exponential fitting parameters, $(A,\alpha,\xi)$ for $U=-1.00$ are $(3.07,0.94,0.001)$.} \label{fig5}
\end{figure}

We take the values for $V_{\text{\normalfont im}}$ in the range of $\Delta (N)$.
In the absence of any impurity, $\Delta$ vanishes exponentially with $N$, at $U=-1.00$. On application of $V_{\text{\normalfont im}} = 0.02$ and $0.10$, $\Delta$ now vanishes as a power law with $N$, as shown in Fig.~\ref{fig5}.
This is clearly distinct from the behavior expected from a gapped topological phase.

\subsection{Comparison with transverse field Ising magnet} \label{sub.TFIM}

Let us now compare the degeneracy properties of the present system to that of a system with conventional symmetry breaking.

We consider a comparison to the degeneracy for a transverse field Ising antiferromagnetic chain realized with Rydberg atoms~\cite{Bernien2017}. Assuming the ordering direction is along the $z$ direction, the ground state of the Ising antiferromagnet is two-fold degenerate between states that have a non-zero on-site magnetization $\langle S^{z}_{j} \rangle \neq 0$, where $S^{z}_{j}$ is the $z$ component of the magnetization at site $j$. The degeneracy in the Ising model is not topological but instead associated with spontaneous breaking of the Ising symmetry ($S^z \rightarrow -S^z$) generated by $S^{x}_{tot} = \sum_j S^{x}_{j}$. However, the symmetry-breaking is qualitatively different from a ferromagnet in the sense that the magnetic order varies in space $\langle S^{z}_{j} \rangle = (-1)^{j} M$, where $M$ is the amplitude of the order parameter. The two ground states of the Ising antiferromagnet are associated with opposite signs of $M$ and are split by a tunneling amplitude that goes to zero exponentially with the length of the system. Similar to the Fermi gas, this degeneracy is not split by a uniform symmetry-breaking Zeeman field in the $z$ direction as long as the Zeeman field varies slowly in space (as long as there is an even number of spins). This is because both states have vanishing total magnetization in the $z$ direction. However, this degeneracy can be seen to be non-topological from the fact that coupling to a magnetic impurity that creates a Zeeman field on a specific site would split the degeneracy by a finite amount. This is analogous to the back-scattering induced splitting in the 1D Fermi gas, although the degeneracy breaking from local impurities is stronger in this example than in the 1D Fermi gas. Therefore, the exponential degeneracy of the ground states in the presence of smooth potentials in the 1D spin-orbit coupled Fermi gas with Zeeman field and attractive interactions cannot be taken to be an indication of a topologically protected degeneracy that can be useful in TQC.

\section{Summary and Conclusion} \label{Discussion}

In summary, we examined the existence and robustness of exponential ground state degeneracy in the presence of local perturbations in a number-conserving 1D Fermi gas with intrinsic attractive interactions, SOC, and a Zeeman field, upon application of a confining parabolic potential in an ultracold-atom system. We found that despite the exponential ground state degeneracy in the clean system, shown in particular by the behavior of the pair binding energy, strictly speaking, the spin-orbit coupled 1D Fermi gas is not in a topological phase because it fails the crucial test of local indistinguishability. Since the exponential ground state degeneracy is not robust against local perturbations, the number-conserving 1D ultracold Fermi gas with attractive interaction is not truly topological and cannot be used for TQC. Our rigorous numerical verification of the theoretical proposal in Ref.~[\onlinecite{Erez_Parabolic}] and the inferences drawn therefrom, will be useful both in the context of designing TQC experiments and for understanding basic aspects of ground state degeneracy and topological properties in 1D interacting Fermi gas systems.

\begin{acknowledgements}\label{Acknowledgement}
{M.K. thanks the Department of Science and Technology (DST), India, for the Ramanujan fellowship and computation facilities provided under DST Project No. SNB/MK/14-15/137. M.K. thanks Prof.~D. Sen and Prof.~J. K. Jain for useful suggestions. J.D.S. was supported by NSF Grant No. DMR1555135 (CAREER) and JQI-NSF-PFC (Grant No. PHY1430094). S.T. acknowledges support from ARO Grant No. W911NF-16-1-0182.}
\end{acknowledgements}

\label{References}



%


\end{document}